\documentclass[9pt,twocolumn,twoside]{pnas-new}
\templatetype{pnasresearcharticle} 
\usepackage{multirow}
\usepackage{algorithm}
\usepackage{algpseudocode}
\usepackage{longtable}
\usepackage{booktabs}
\usepackage{tabularx}
\usepackage{array}
\usepackage{rotating}

\begin{document}

\title{Upstream flow geometries can be uniquely learnt from single-point turbulence signatures}

\author[a]{Mukesh Karunanethy}
\author[b]{Raghunathan Rengaswamy}
\author[a,1]{Mahesh V Panchagnula}

\affil[a]{Department of Applied Mechanics and Biomedical Engineering, Indian Institute of Technology Madras, Chennai, Tamil Nadu, India}
\affil[b]{Department of Chemical Engineering, Indian Institute of Technology Madras, Chennai, Tamil Nadu, India}

\leadauthor{Karunanethy}

\significancestatement{
Turbulence has long been characterized by reducing a vector of velocity-time series data to scalar metrics from probability distributions (e.g., standard deviation). We show that a much richer vector of invariant measures, especially the temporal autocorrelation coefficients are robust, informative and discriminatory. We test the hypothesis that the near-field turbulent flow downstream sudden contractions created using orifices of different shapes is uniquely related to the orifice shape. We show that the invariant scalar measures carry signatures sufficient for a machine learning algorithm to uniquely identify the shape of the orifice. While the chosen illustrative problem is from fluid dynamics, the demonstration that ML can learn from a vector of invariant measures and robustly identify systems is of great broader interest.}

\authorcontributions{Author contributions:  M.K., R.R., and M.V.P. designed and performed research; M.K. recorded and analysed data; M.K., R.R., and M.V.P. wrote the paper.}
\authordeclaration{The authors declare no competing interests.}
\correspondingauthor{\textsuperscript{1}To whom correspondence should be addressed. Email: mvp@iitm.ac.in}


\begin{abstract}
We test the hypothesis that the microscopic temporal structure of near-field turbulence downstream of a sudden contraction contains geometry-identifiable information pertaining to the shape of the upstream obstruction. We measure a set of spatially sparse velocity time-series data downstream of differently-shaped orifices. We then train random forest multiclass classifier models on a vector of invariants derived from this time-series. We test the above hypothesis with 25 somewhat similar orifice shapes to push the model to its extreme limits. Remarkably, the algorithm was able to identify the orifice shape with 100$\bf{\%}$ accuracy and 100$\bf{\%}$ precision. This outcome is enabled by the uniqueness in the downstream temporal evolution of turbulence structures in the flow past orifices, combined with the random forests' ability to learn subtle yet discerning features in the turbulence microstructure. We are also able to explain the underlying flow physics that enables such classification by listing the invariant measures in the order of increasing information entropy. We show that the temporal autocorrelation coefficients of the time-series are most sensitive to orifice shape and are therefore informative. The ability to identify changes in system geometry without the need for physical disassembly offers tremendous potential for flow control and system identification. Furthermore, the proposed approach could potentially have significant applications in other unrelated fields as well, by deploying the core methodology of training random forest classifiers on vectors of invariant measures obtained from time-series data.
\end{abstract}

\dates{This manuscript was compiled on \today}

\maketitle
\ifthenelse{\boolean{shortarticle}}{\ifthenelse{\boolean{singlecolumn}}{\abscontentformatted}{\abscontent}}{}

\firstpage[9]{3}

\dropcap{T}he convergence of fluid mechanics and machine learning represents a rapidly growing interdisciplinary field with immense potential for a deeper understanding of complex flow phenomena. This manuscript discusses Machine Learning (ML) models developed to study the uniqueness of flow complexity in turbulent flows. We hypothesize and show that these unique flow signatures, captured in velocity time series data, can be used to accurately differentiate between various upstream geometrical features using ML classifier models. The flow through a circular pipe with blunt orifices is a well-studied canonical fluid dynamics problem and is used for flow measurement extensively. We use this as the problem to illustrate the potential of exploiting the microscale turbulent flow features.

Since the pioneering studies by Osborne Reynolds \cite{Reynolds1883}, researchers have studied turbulent wall-bounded flows through pipes and channels with obstructions. Such flows result in confined jets, which are typically expanding jets within pipes. Such flows are characterized by a potential core, a mixing region and a recirculation zone \cite{Curtet1958, Rajaratnam1976confinedjets}. The flow separation or recirculation of such confined axisymmetric jets was investigated experimentally and analytically by Exley et al. \cite{Exley1971}. The characteristics of turbulent confined jets were studied numerically by Kandakure et al. \cite{Kandakure2008}, who also provides a comparison of several flow parameters from the literature. These studies highlighted the significant impact of geometrical parameters on jet behaviour, including entrainment rates and turbulent kinetic energy. While these studies pointed to a set of fundamental flow features of confined jets, the objective of this study is to go beyond and show that the near-field turbulence structure contains orifice shape-linked information. As we will show, downstream flow features is uniquely linked to the upstream orifice shape.

One of the earliest reports where Machine Learning met fluid dynamics is due to Muller et al. \cite{Muller1999}, where they demonstrated the application of machine learning algorithms to flow modelling and optimization. Zhang et al. \cite{Zhang2015} explored the application of machine learning for data-driven turbulence modelling, by introducing a multi-scale Gaussian process regression method that outperformed conventional methods in reconstructing spatially distributed functions from high-dimensional, noisy data. Beck et al. \cite{Beck2021} provided a comprehensive review of data-driven turbulence modelling, highlighting the challenges and potential of integrating machine learning methods alongside large-eddy simulations and other related CFD methods. These studies highlight the potential for machine learning as an enabling tool in data-driven turbulence modelling. Deep neural networks play a critically enabling role in modelling complex flows \cite{Kutz2017}. Karniadakis et al. \cite{Karniadakis2021} explored the intersection of machine learning and physics using Physics-Informed Machine Learning (PIML). By seamlessly integrating empirical data alongside mathematical physics models, PIML aims to improve performance on tasks involving physical mechanisms. Machine learning has been pivotal in several studies involving experimental data. In the present work, machine learning is utilised for the classification of time series features attributed to different orifice geometries. Given a set of extracted invariant time series features, supervised machine learning models known as classifiers are built, which effectively demarcate the difference between different flow fields in the invariant feature space. Two widely used classification models/algorithms in the literature are support vector machines (SVM) \cite{Schlkopf2018} and random forests (RF) \cite{Breiman2001}. We choose the latter for the present work.

Machine learning methodologies applied to computational and experimental fluid dynamics have led to innovative approaches for analyzing and interpreting data. Zhou et al. \cite{Zhou2020} developed an artificial intelligence-based control system for a turbulent jet that optimizes mixing rates by discovering and implementing advanced forcing strategies. Li et al. \cite{Li2022} presented a comprehensive review of the application of machine learning on fluid mechanics datasets propounding an in-depth understanding of the underlying flow mechanisms. Such understanding will assist in flow modelling and active flow control in real-time systems \cite{Fan2020}. Several other studies \cite{Brunton2020b, Brunton2020a, Brunton2021} have focused on discussing how machine learning algorithms and data-driven approaches help in solving problems in fluid dynamics. High-fidelity computational fluid dynamics methods, such as Direct Numerical Simulation (DNS) and Large-Eddy Simulation (LES), allow us to apply machine learning to spatial features. For example, researchers have built models to identify turbulent and non-turbulent regions within a flow field \cite{Li2020turbulentregion}. In order to conduct such studies experimentally, expensive high-resolution optical measurement techniques like 2D/3D Particle Image Velocimetry (PIV) would be required. High-resolution spatio-temporal reconstruction of turbulent flows using supervised machine learning techniques have been explored in the past \cite{Fukami2019, Fukami2020, Fukami2023}. Identifying regions of such flow fields, based on a single-point velocity data is an interesting challenge that has not been attempted in the literature. This work proposes to build machine learning classifier models that can identify the uniqueness of a flow given a single point velocity time series measurement. 

Hotwire Anemometry is a great tool to derive time-resolved voltage signals related to the flow velocity at a single measurement point in space, owing to its high temporal resolution. However, this raw velocity (or voltage) time series is usually not well-suited as input data for most machine learning algorithms \cite{Faouzi2022}. Extracting relevant invariant features from experimental time series data is critical in efficiently analyzing flow behaviour and building robust machine learning models. In the context of this work, these features will be used to train classifier models. Time series parameters have already been shown to be good classifying tools in many engineering systems. Acharya et al. \cite{Acharya2021} showed the significance of time series features in classifying size-velocity data of sprays measured using a phase Doppler particle analyser. Godavarthi et al. \cite{Godavarthi2019} demonstrated a similar classification of spray data based on the time signal features extracted using multifractal techniques. A study by Oriona et al. \cite{Oriona2021} emphasises the importance of time series parameters in achieving accurate classification results. The following are some of relevant time series feature extraction tools available in the literature: $(i)$ FATS (Feature Analysis for Time Series) \cite{nun2015fats}, $(ii)$ \textit{tsfresh} (Time Series FeatuRe Extraction on the basis of Scalable Hypothesis tests) \cite{CHRIST2018}, $(iii)$ HCTSA \cite{Fulcher2017}, and $(iv)$ TSFEL (Time Series Feature Extraction Library) \cite{BARANDAS2020}. These packages make use of several time series analysis techniques to generate automated features.

In summary, this study introduces a revolutionary hypothetical question: ``Can a flow geometry be identified uniquely and solely from a sparse set of single-point velocity time series measurements?'' Answering this question holds the potential to enable various applications in flow field reconstruction, fluid controls, and personalised medicine. The use of hot wire anemometer and, more broadly, turbulence measurements for identifying obstructions in a flow has not been attempted in the literature. We hypothesize that different patterns of disturbances imparted to the flow by different orifice geometries can be inferred from a single measured time series. In this study, we will conclusively show that the flow field carries sufficient information to identify different flow systems and orifice geometries. While the example problem chosen is fluid dynamic in nature, the methodology of applying ML models as classification tools to invariant features extracted from time series data holds great promise to fields beyond fluid dynamics. We wish to open that possibility by showcasing the nuanced and subtle nature of the ML models in their classification abilities.

\section*{Materials and Methods}

The experimental setup consisted of a mass flow controller, a settling chamber, and a pipe system with interchangeable orifice plates (see \textit{SI Appendix}, Fig. S1), enabling time-series measurements downstream of and in the near field of $25$ different orifice geometries (see \textit{SI Appendix}, Table S8). Table \ref{tab:geometry_id_mapping} shows the list of $25$ orifice geometries and their respective identities. Measurements were made using a hot wire anemometer at nine locations, labelled as `A' through `I', at a chosen distance downstream of the orifice, with consistent flow conditions maintained across measurements. A detailed list of datasets for various flow characteristics, orifice geometries, and test configurations is provided in \textit{SI Appendix}, Table S1.

\begin{SCfigure*}[\sidecaptionrelwidth][t!]
\centering
\includegraphics[scale=0.15]{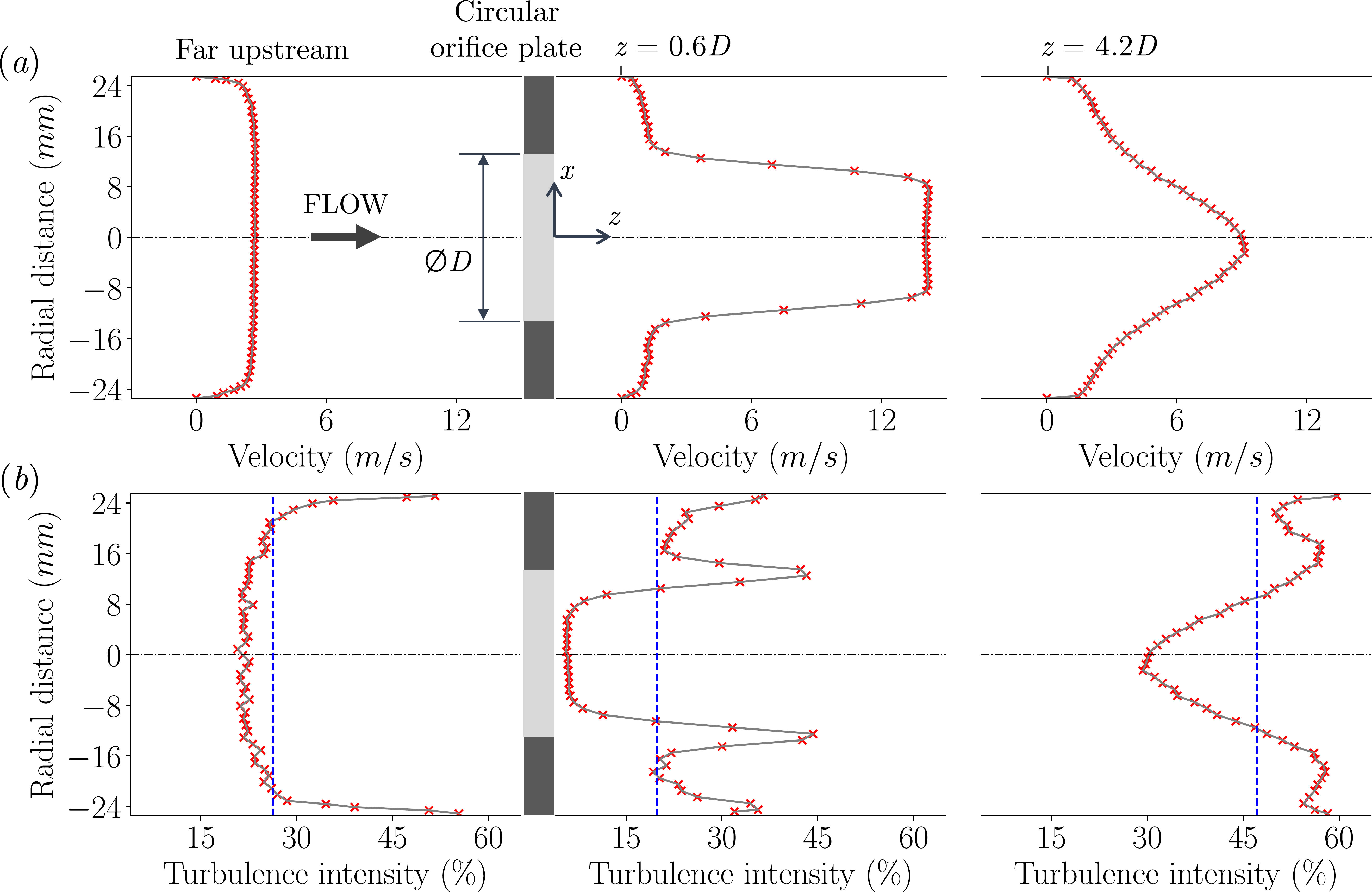}
\caption{$(a)$ Measured velocity profiles at distances far-upstream, $z=0.6D$ downstream, and $z=4.2D$ downstream of a circular orifice plate. $(b)$ Turbulence intensity profiles at distances far-upstream, $z=0.6D$ downstream, and $z=4.2D$ downstream of the orifice plate. The red ({\color{red}$\times$}) markers in $(a)$ and $(b)$ denote the measured mean velocity values at each point in the cross-section. The blue dotted line in $(b)$ represents the mean turbulence intensity value at each cross-section. All the measurements were made for a Reynolds number of $2.37\times 10^4$. The flow direction is from left to right. Diameter of the orifice plate, $\varnothing D=26.8mm$.}
\label{fig:profiles}
\end{SCfigure*}

\subsection*{Flow conditions and velocity profiles}

Initially, measurements were performed to construct the mean velocity and turbulence intensity profiles at different cross-sections: $(i)$ far upstream of the orifice plate, $(ii)$ at $z=0.6D$ (near field), and $(iii)$ at $z=4.2D$ (far field) downstream of the orifice plate. $D$, here, is the diameter of a circular orifice ($=26.8\times10^{-3}m$), and the distances were measured in the positive $z$ direction from the plate. Velocity time series of $5$ seconds were measured at $52$ different locations across each cross-section. This was achieved by traversing the hot wire probe (along $x$ direction with reference to \textit{SI Appendix}, Fig. S1) from one diametric end to the other end within the pipe. Note that the velocity at the wall is assumed to be $0m/s$ due to the no-slip condition. These measurements were made for a Reynolds number of $2.37\times 10^4$ corresponding to a mass flow rate of $300slpm$. Care was taken to ensure that both the upstream mean velocity and turbulence intensity profiles, as depicted in Fig. \ref{fig:profiles}$(a)$ and Fig. \ref{fig:profiles}$(b)$, were unchanged for all the experiments conducted during this study. This ensures that the ML algorithm is not relying on spurious features associated with the initial condition of the flow. In addition, by interrogating the velocity profiles, one can observe the manner in which the flow evolves spatially and changes behaviour as it progresses through the system. The profile far upstream is a fully developed turbulent velocity distribution typically observed in a flow inside a pipe of a circular cross-section. As the flow interacts with the orifice, shear is imparted and the flow characteristics change. This shearing action is hypothesized to be unique for each orifice geometry. Furthermore, the turbulence intensity ($TI$) at the above-mentioned cross-sections is visualised in Fig. \ref{fig:profiles}$(b)$. They provide information regarding the degree of turbulence present in the flow. The turbulence intensity is computed as the percentage of root-mean-square velocity to the mean velocity. It serves as a key parameter for assessing the turbulent nature of the flow. By visualizing the turbulence intensity distribution at different cross-sections, we can identify regions of high turbulence activity and study the effect of various flow conditions or configurations. In both velocity and turbulence intensity profiles, the red markers denote the measured mean values at each point in the cross-section.

Additionally, the blue dotted line in Fig. \ref{fig:profiles}$(b)$ represents the mean turbulence intensity value at each cross-section, summarizing the overall turbulence levels at each cross-section. A variation in turbulence intensity ($TI$) levels is observed as the flow progresses from an upstream location and past the orifice plate as described below:

\begin{itemize}
    \item Far upstream, the mean $TI$ is around $26\%$, indicating moderate turbulence levels and remained nearly constant for all experiments.
    \item At $z=0.6D$, the mean $TI$ drops to around $20\%$ as expected and remains relatively moderate. This reduction is due to the sudden contraction of the area, accelerating the flow near the axis, which causes a vena contracta effect. It is well known that this creates a stabilising or damping effect in the flow, leading to decreased turbulence levels. In the shear layer (downstream of the edge of the orifice), as expected again, a high level of turbulence, roughly around $45\%$ is observed. This falls in the region of interest for the current work.
    \item At $z=4.2D$, the mean $TI$ significantly increases to around $47\%$, indicating a substantial increase in the turbulence levels further downstream. Again, as expected, the effect of the shear layer is observed to diffuse in the radial direction, causing the increased $TI$.
\end{itemize}

\subsection*{Time signal} \label{sec:time_signal}

\begin{SCfigure*}[\sidecaptionrelwidth][t!]
\centering
\includegraphics[scale=0.35]{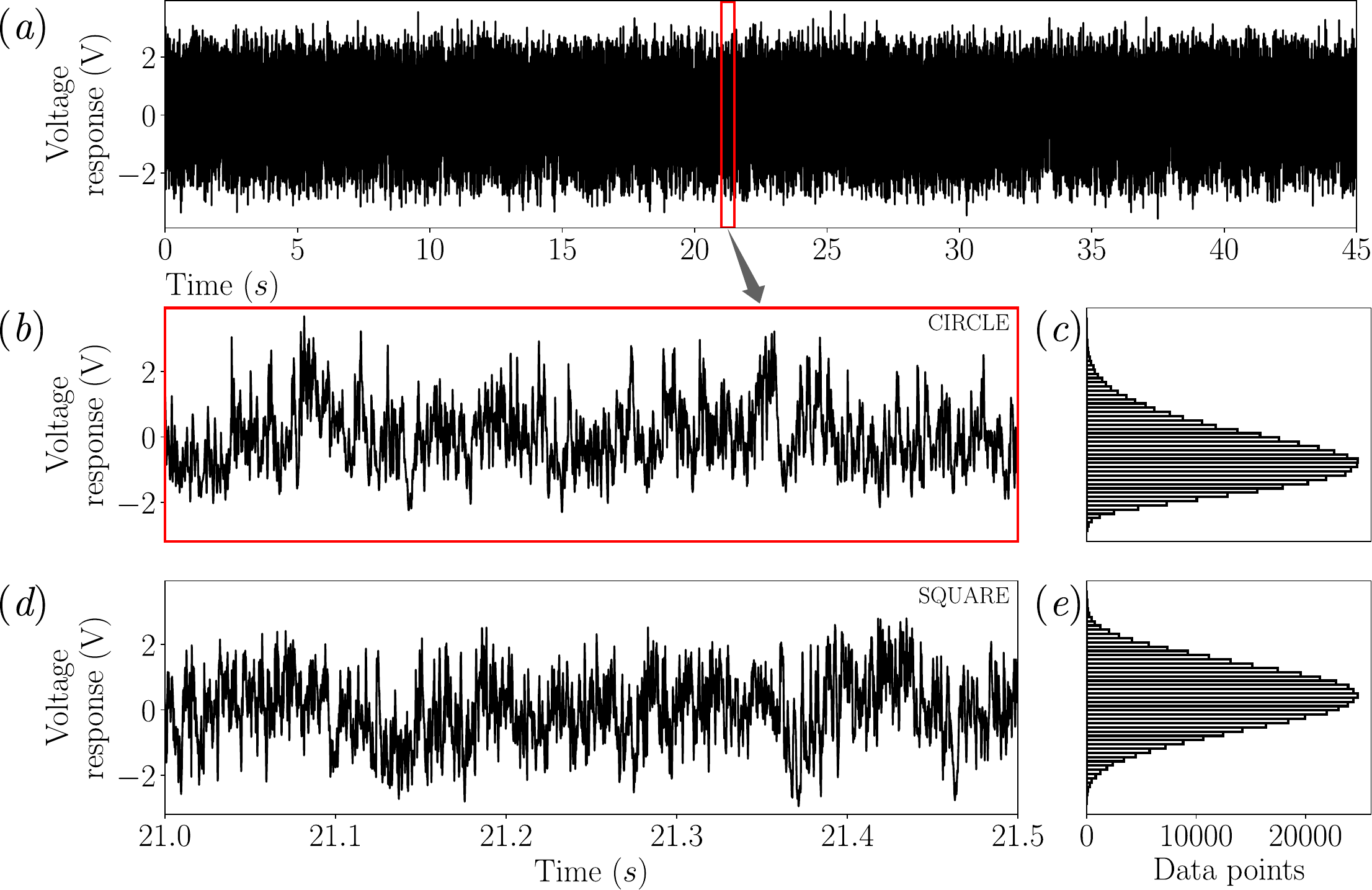}
\caption{Comparison of time series signals measured downstream of the circular and square orifices (near location `H' in \textit{SI Appendix}, Fig. S1). $(a)$ A typical time signal of $45$ seconds corresponding to a circular orifice, recorded at a sampling rate of $10kHz$. $(b)$ A stretched-out visualisation of a $0.5$ second time window corresponding to the segment within the red box of $(a)$. $(d)$ Similar zoomed-in time signal view for a square orifice geometry. $(c)$ and $(e)$ are the histograms showing the distribution of all $450,000$ data points of the signals corresponding to circular and square geometries, respectively. Note that these signals are normalised using their mean and standard deviation.}
\label{fig:time_series_circle_square}
\end{SCfigure*}

A typical time signal recorded for $45$ seconds at a location downstream of an edge of a circular orifice (near location `H' in \textit{SI Appendix}, Fig. S1) is shown in Fig. $\ref{fig:time_series_circle_square}(a)$. Fig. $\ref{fig:time_series_circle_square}(b)$ shows a zoomed-in view of the signal in Fig. $\ref{fig:time_series_circle_square}(a)$ between times of $21.0$ and $21.5$ seconds. All the signals were recorded at a sampling rate of $10kHz$. This effectively yields $450,000$ data points in one recorded signal block. Fig. $\ref{fig:time_series_circle_square}(d)$ shows a similar zoomed-in view of a time signal recorded in the near field downstream of an edge of a square orifice. Note that these signals are normalised using their mean and standard deviation. The histograms shown in Fig. $\ref{fig:time_series_circle_square}(c)$ and Fig. $\ref{fig:time_series_circle_square}(e)$ are the distribution of all data points of the signals corresponding to circle and square geometries, respectively. While comparing these signals, one could be led to observe the presence of generally lower frequency content downstream of the circular orifice (Fig. $\ref{fig:time_series_circle_square}(b)$) in comparison to the signal downstream of the square orifice (Fig. $\ref{fig:time_series_circle_square}(d)$). The distributions also show a slight skewness for the circle (Fig. $\ref{fig:time_series_circle_square}(c)$), whereas it is reasonably symmetric for the square orifice (Fig. $\ref{fig:time_series_circle_square}(e)$). While these visual cues are arguably there, what matters is whether an ML algorithm can conclusively and statistically identify these differences. In general, it is interesting to observe that within a short time window of $0.5$s, there exists significant fluctuations and provides the motivation to study how unique the features pertaining to signals corresponding to each orifice geometry would vary.

\textit{SI Appendix} details the pre-processing steps and classification models used in this study. The time series were segmented, and z-score normalization was applied to generate representative samples for model training. Time series features were extracted using the \textit{tsfresh} tool \cite{CHRIST2018}. Subsequently, all the low-variance as well as highly correlated features were removed to reduce problem dimensionality. Random forest classifiers were then trained using a Bayesian Search Cross-Validation strategy, optimizing hyperparameters (see \textit{SI Appendix}, Table S3 for the list of hyperparameters) to ensure robust model performance across multiple classes.

\subsection*{Random forest classifier}

A multi-class random forest classifier model was trained to classify the $25$ orifices. To eliminate any potential measurement bias, the list of orifices was randomly shuffled using a random number generator prior to training the model. The orifice identity information is shown in table \ref{tab:geometry_id_mapping}. The trained model is applied to a given test data (pertaining to a test orifice); it makes predictions on all the data points within the test data. The model assigns each data point a label/identity.  Algorithm \ref{alg:CA} describes the methodology for orifice confirmation. The classifier's output can be interpreted as the model prediction probability, which is a vector denoted by $\Phi_i$, for `Orifice $i$'. This vector's elements $a_1, a_2, \ldots, a_n$ represent the probability corresponding to each class label ($n$ is the total number of orifices). For a given orifice's test data, the orifice confirmation was then performed by setting a threshold for the probability. Here, a threshold of $50\%$ was set for defining the minimum confidence of confirmation. This means that if the probability corresponding to the `Orifice $i$' is greater than $0.5$, the algorithm confirms the orifice's identity; otherwise, it does not confirm it.

\begin{algorithm*}[t!]
\caption{Pseudocode of an orifice confirmation algorithm based on multi-class classifier approach}
\label{alg:CA}
\begin{algorithmic}[1]
\Function{ConfirmationAlgorithm}{$orifice\_id,\, test\_data,\,$ MODEL(), $threshold=0.5$} \textbf{returns} a message
    \State $\Phi \gets$ MODEL($test\_data$) \Comment{Make predictions using the multi-class classifier}
    \State $p \gets \Phi[orifice\_id]$  \Comment{Get the confirmation probability based on the $orifice\_id$}
    \If{$p > threshold$}
        \State \textbf{return} ``Orifice confirmed.''
    \Else
        \State \textbf{return} ``Orifice not confirmed.''
    \EndIf
\EndFunction
\end{algorithmic}
\end{algorithm*}

The orifice identification algorithm focuses on identifying the orifice in the absence of prior knowledge. As observed from algorithm \ref{alg:IA2}, the output of the random forest model was a prediction probability vector, $\Phi_i$, of size $(1,n)$. The identified orifice from this algorithm will then be the orifice corresponding to the maximum probability in the vector $\Phi_i$. If multiple classes have the same maximum probability, the algorithm does not identify a unique orifice. For the rest of the manuscript, we will only present orifice shape identification results since identification demands that the algorithm {\em{identify}} the orifice with no prior information.

\begin{algorithm*}[t!]
\caption{Pseudocode of an orifice identification algorithm based on the multi-class classifier approach}
\label{alg:IA2}
\begin{algorithmic}[1]
\Function{IdentificationAlgorithm}{$test\_data,$ MODEL()} \textbf{returns} orifice identity
    \State $\Phi \gets$ MODEL($test\_data$) \Comment{Make predictions using the multi-class classifier}
    \State \textbf{return} $\operatorname{argmax}(\Phi)$ \Comment{Identified orifice is the orifice corresponding to maximum probability}
\EndFunction
\end{algorithmic}
\end{algorithm*}

\section*{Results}

\subsection*{Orifice identification results}

Features were extracted from all the velocity time series datasets listed out in \textit{SI Appendix}, Table S1. Recall that the recorded time signals were segmented and normalised before feature extraction, as described in earlier sections. The performance of random forest models (binary classifier and multi-class classifier approaches) was evaluated using the test dataset. A comparison of the algorithms and results from the binary classifier approach has been discussed in the \textit{SI Appendix}. For brevity, only the results from the multi-class classifier are presented herein.

\begin{table}[t!]
\centering
\caption{A list of orifice geometry names and their respective orifice identities (ID). Note that the list of geometries was not sorted in any geometric order. The orifice ID in the first column shows the randomly shuffled order in which the geometries were arranged for model training and algorithm testing}
\label{tab:geometry_id_mapping}
\begin{tabular}{c l l}
        \toprule
        Orifice ID & Geometry \\
        \midrule
        0 & Right-angled Triangle \\
        1 & Four-pointed Star \\
        2 & Square \\
        3 & Pentagon \\
        4 & Kite \\
        5 & Obtuse-angled Triangle \\
        6 & Circle \\
        7 & Ellipse \\
        8 & Rectangle \\
        9 & Acute-angled Triangle \\
        10 & Equilateral Triangle \\
        11 & Parallelogram \\
        12 & Semi Circle \\
        13 & Heart \\
        14 & Quadrant \\
        15 & Rhombus \\
        16 & Six Pointed Star \\
        17 & Hexagon \\
        18 & Trapezium \\
        19 & Heptagon \\
        20 & Seven-pointed Star \\
        21 & Five-pointed Star \\
        22 & Isosceles Triangle \\
        23 & Octagon \\
        24 & Eight-pointed Star \\
        \bottomrule
    \end{tabular}
\end{table}

\begin{SCfigure*}[\sidecaptionrelwidth][t!]
\centering
\includegraphics[scale=0.45]{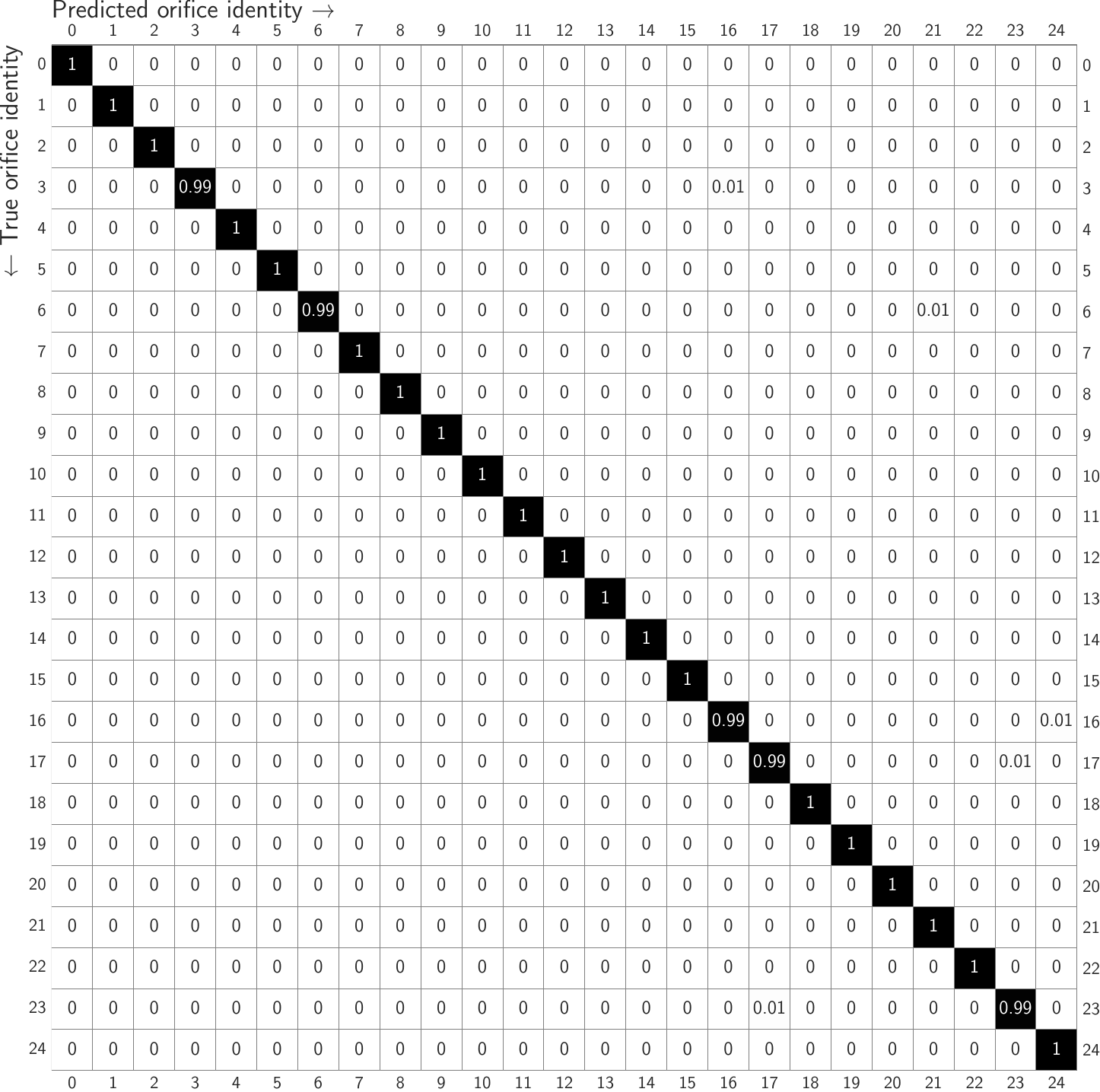}
\caption{The confusion matrix $(\mathbf{K})$ contains the probability of prediction for each possible orifice identity. Each row represents a tested orifice identity, while each column represents one of the orifice identities for which the model has been trained. Since the model was trained for all $25$ geometries, and all those geometries were tested for performance, the shape of this matrix is $25\times25$. In each row, the cell with the highest probability has been marked black.}
\label{fig:confusion_matrix}
\end{SCfigure*}

The overall outcomes from a multi-class random forest classifier can be visualised in a confusion matrix, $\mathbf{K}$, an example shown in Fig. \ref{fig:confusion_matrix}.  All those geometries were tested for performance, and hence, the shape of this matrix is $25\times25$. $\mathbf{K}$ can be visualized as a stack of prediction probability vectors $(\Phi_i)$ for all the $n$ available orifice geometries. The columns of the confusion matrix would then denote the predicted orifice's label/identity, and the rows denote the actual orifice's label/identity. The identified orifice geometry corresponds to the cell with the highest prediction probability. Those cells are highlighted in each row of Fig. \ref{fig:confusion_matrix}. From an orifice confirmation perspective (diagonal of the confusion matrix), all orifices passed well over the threshold of $0.5$ and were confirmed correctly.

The following performance metrics were used to evaluate the orifice geometry confirmation and identification algorithms: True Confirmation Rate $(TCR)$, which measures the algorithm's ability to correctly confirm all known orifice geometries. $TCR$ in the current experiment is $100\%$, indicating that the algorithm is able to correctly confirm all the known orifice geometries; Accuracy $(A)$, which quantifies the percentage of correctly identified orifice geometries. The experiment also shows an accuracy of $100\%$, indicating that the algorithm shows no false negatives; Precision $(P)$, which assesses the proportion of correctly identified orifice geometries among all identified ones. $P$ is also $100\%$, indicating maximum precision in orifice identification. For detailed definitions of the performance metrics, see \textit{SI Appendix}.

The above reported results of $TCR$, Accuracy and Precision being $100\%$, utilized the entire data set from all nine probe locations (`A'$-$`I') for orifice identification. We will now investigate if any one point or any set of points from among the nine probe locations, `A'$-$`I', are more discriminatory than others. Towards that end, we run the Random Forest multi-class classifier with a limited data set corresponding to one or more of the probe locations. Table \ref{tab:performance_at_locations} presents the performance metrics of the orifice identification algorithms by selecting one or more of the nine probing locations downstream of the orifice plate, shown in \textit{SI Appendix}, Fig. S1. It has been observed that the $TCR$ for any one single point considered alone is not satisfactory. However, considering groups of four points, either (`B', `D', `G', `H') or (`A', `E', `F', `I') is as good as considering all nine points `A'$-$`I'. Interestingly, the precision was $100\%$ at all locations, suggesting that the identified orifice geometries based even on single-point data are not false. Finally, as expected, the probe location `C' exhibits the lowest discriminatory capability compared to all other locations, on both $TCR$ and accuracy measures. This result aligns with general fluid dynamic intuition that for all geometries, the features near the pipe axis are in the potential core of the jet and hence are least informative.

\begin{table}[t!]
\centering
\caption{Performance metrics of the orifice confirmation and identification algorithms for a database of $25$ geometries. `TCR' stands for true confirmation rate. The results are based on $100$ realisations of train-test shuffling}
\label{tab:performance_at_locations}
\begin{tabular}{c c c c}
        \toprule
        Location & $TCR$ (\%) & $A$ (\%) & $P$ (\%) \\
        \midrule
        A & $60 \pm 3.7$ & $60 \pm 3.7$ & $100 \pm 0$ \\
        B & $78 \pm 3.2$ & $78 \pm 3.2$ & $100 \pm 0$ \\
        C & $33 \pm 2.8$ & $33 \pm 2.8$ & $100 \pm 0$ \\
        D & $91 \pm 3.0$ & $91 \pm 3.0$ & $100 \pm 0$ \\
        E & $40 \pm 2.5$ & $40 \pm 2.5$ & $100 \pm 0$ \\
        F & $47 \pm 3.8$ & $47 \pm 3.8$ & $100 \pm 0$ \\
        G & $91 \pm 2.4$ & $91 \pm 2.4$ & $100 \pm 0$ \\
        H & $74 \pm 3.0$ & $74 \pm 3.0$ & $100 \pm 0$ \\
        I & $43 \pm 3.9$ & $43 \pm 3.9$ & $100 \pm 0$ \\
        B D G H & $100 \pm 0$ & $100 \pm 0$ & $100 \pm 0$ \\
        A E F I & $100 \pm 0$ & $100 \pm 0$ & $100 \pm 0$ \\
        A$-$I & $100 \pm 0$ & $100 \pm 0$ & $100 \pm 0$ \\
        \bottomrule
        \end{tabular}
\end{table}

Data from locations `D' and `G' exhibited the best performance among all locations. Following that, the locations `A', `B' and `H' demonstrated moderate performances with $TCR$ and accuracy ranging from around $60\%$ to $80\%$. Locations `E', `F' and `I' exhibit lower $TCR$ and accuracy, varying between $40\%$ to $50\%$. Apart from the analysis of individual locations, combinations of locations were analysed: $(i)$ `B', `D', `G', `H' combined, $(ii)$ `A', `E', `F', `I' combined, and $(iii)$ All locations `A'$-$`I' combined. The confirmation and identification results have been tabulated in the last three rows of table \ref{tab:performance_at_locations}. The algorithm was observed to work perfectly with $100\%$ true confirmation rate, $100\%$ identification accuracy and precision for these three cases.

\subsection*{Repeatability and robustness of the algorithm}

The alignment of the system was maintained unchanged throughout the data recording of all the $25$ geometries. Throughout this manuscript, the term \textit{alignment} would refer to the relative positioning of the hot wire traverse mechanism with respect to the pipe. The traverse should be made in the $x$, $y$ and $z$ directions (orthogonal axes shown in \textit{SI Appendix}, Fig. S1) with the $z$ axis exactly aligning with the axis of the pipe. Minor misalignment in any of these directions or a combination of these directions could lead to asymmetry in the measurements. This was a major challenge in the experimental setup that could affect the repeatability of the results. It is important to verify that the system identification is robust and invariant to minor system alignment issues.

Rigorous tests were performed to ensure the robustness and repeatability of the random forest classifier, involving multiple system alignments on three different representative orifice shapes: circle, square and equilateral triangle and over multiple days. The results showed consistently high prediction accuracy across trials despite slight variations due to alignment imperfections (see \textit{SI Appendix}, Table S4 for details). Three different CTAs and two different hot film probes, as listed in \textit{SI Appendix}, Table S2 were also used during the repeatability study. From this study, we believe that the conclusion that the velocity time series in the shear region contains system identifiable information, is robust.

The classifier performance was also evaluated for various Reynolds numbers ($Re$), where the model achieved consistently high accuracy, particularly at $Re=2.37\times 10^4$ and $Re=2.77\times 10^4$ (see \textit{SI Appendix}, Table S5 and \textit{SI Appendix}, Table S6 for details). Models built on data gathered from these different $Re$ can also be used across a wider range of $Re$ values.

Additionally, an analysis of the classifier performance as a function of the axial location of the probe locations, showed that the classifier’s accuracy, as expected,  decreases as the distance from the orifice increases. The accuracy was observed to be nearly $100\%$ at $z=0.6D$ and $97\%$ at $z=1.9D$ followed by $78\%$ at $z=4.2D$. This implies that the orifice-specific information is gradually lost, with the strongest classification occurring in the near field at $z=0.6D$ and accuracy decreasing beyond about two orifice diameters (see \textit{SI Appendix}, Table S7 for details).

\begin{figure}[t!]
\centering
\includegraphics[width=.98\linewidth]{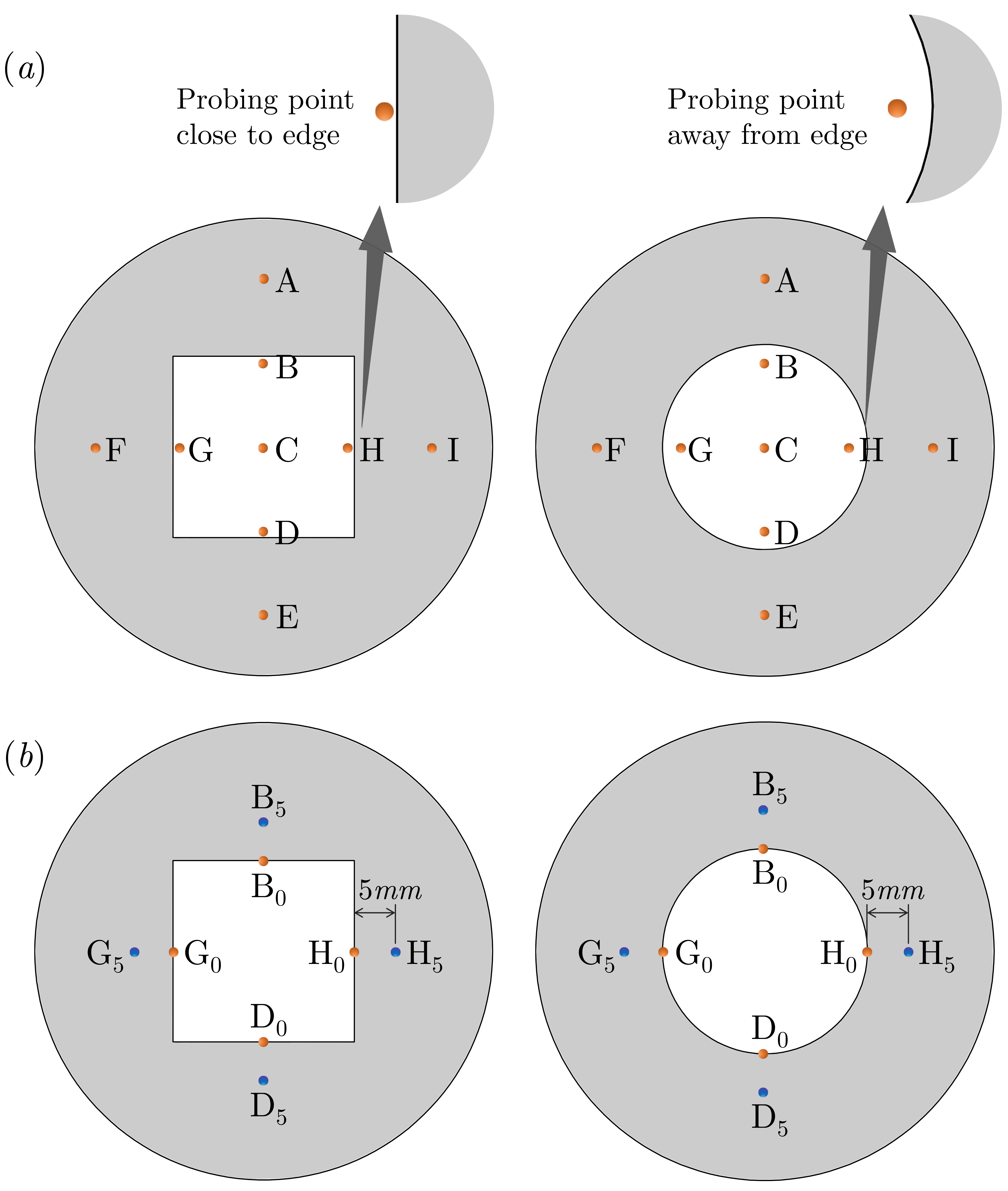}
\caption{$(a)$ Observation showing the difference in distance of the hot wire probing location from the edges of the square and circular geometry orifice plates. The orange dots in $(a)$ represent the $9$ probing points, labelled `A', `B', `C', `D', `E', `F', `G', `H' and `I'. The results discussed in previous sections pertain to measurements from these $9$ locations. $(b)$ New locations assigned for hot wire measurements to investigate the edge effects on the performance of classifier models. The orange points in $(b)$ represent the probing locations immediately downstream of the orifice edge, labelled `B\textsubscript{0}', `H\textsubscript{0}', `D\textsubscript{0}', and `G\textsubscript{0}', and the blue points represent probing locations $5mm$ away from each edge location, towards the wall, labelled `B\textsubscript{5}', `H\textsubscript{5}', `D\textsubscript{5}', and `G\textsubscript{5}'.}
\label{fig:edge_data_points}
\end{figure}

Based on the observations presented, one might question whether the distance of the probing points from the orifice edges becomes a deciding factor for classification at a given distance from the orifice. Since the probing locations are located at fixed distances from the center-line of the system, it is possible that the uniqueness in the flow characteristics could arise out of the fact that the probing was performed at differing distances from the edges of the various orifices. In order to illustrate the point, the observable distances of these points on the square- and circle-shaped orifices are visualised in Fig. $\ref{fig:edge_data_points}(a)$. In order to conclusively preempt the question, a set of measurements were made downstream of the four edges (near each of the `B', `H', `D', and `G' locations, labelled as `B\textsubscript{0}', `H\textsubscript{0}', `D\textsubscript{0}', and `G\textsubscript{0}', respectively) and at exactly $5mm$ away from each edge towards the wall of the pipe, as shown in Fig. $\ref{fig:edge_data_points}(b)$. These points were labelled as `B\textsubscript{5}', `H\textsubscript{5}', `D\textsubscript{5}', and `G\textsubscript{5}' for clarity. By this approach, we could know whether the time series features that were responsible for the model performance are invariant to small changes in the position of the probe location point near the edge. The probing locations were manually aligned by placing the probes aligned to the edge and then translating the probe $5mm$ in the required direction. The training and test data were also recorded on two different days to ensure robustness. The training data includes $5$ signals of $45s$. The testing data consists of two $45s$ signals. All the samples were recorded at $10kHz$. These were the datasets shown on rows 13 and 14 of \textit{SI Appendix}, Table S1. Binary random forest classifier models were built for the two cases (edge data and data from $5mm$ away) for circular and square orifices. For both cases, models were observed to achieve $100\%$ accuracy. Therefore, it can be concluded that the distance from the edge is not the sole reason for the classification and that the classification is based on fluid dynamic information in the time series. Sampling one component of a 3-dimensional velocity profile downstream (even sparsely) is sufficient for a machine learning model to learn discriminative features for classification.

\section*{Discussion}

\subsection*{Essential features and flow-physics-based arguments}

We have provided conclusive evidence that identifying orifice geometry based on velocity time series measurements is feasible. To understand the underlying physics and explain the classifier performance from physics, it is important to understand the physical significance of the important time series features, which are responsible for robust classification. Since random forest classifiers were used, the importance of features can be quantified based on the information impurity, for a given training dataset. Important features were extracted from random forest models trained with a dataset from each probing location `A'-`I' separately. A union of features were obtained from all these features. These features are then investigated to understand their physical meaning in the context of the flow past orifices. The features and their physics-based description are listed hereunder.

\begin{enumerate}

    \item Second coefficient of the autoregressive $AR(r)$ model with order parameter $r = 10$: The parameter $r$ is the maximum lag of the autoregressive process. The value of a time series $X_t$ can be related to $i$ time steps before the current time using the following equation.
    \begin{equation}
        X_t=\varphi_0+\sum_{i=1}^r \varphi_i X_{t-i}+\varepsilon_t
    \end{equation}
    Here, $\varphi_0$ is the constant or intercept term. It represents the baseline or the mean value of the time series, $\varphi_i$ are the autoregressive coefficients ($i=1,2,\ldots,r$), and $\varepsilon_t$ is the error term that represents the difference between the observed value and the predicted value based on the autoregressive model.\\
    The $AR$ model generally predicts future behaviour based on past data. The importance of the autoregressive coefficients shows that there is some correlation between successive values in the time series for most of the orifice geometries. This indicates that past values of the velocity time series contain useful information for predicting future values. Such an observation indicates that the flow downstream of different orifices exhibit different levels of temporal dependencies and persistence.

    \item Fifth coefficient of the autoregressive $AR(r)$ model with order parameter $r = 10$.

    \item Fourth coefficient of the autoregressive $AR(r)$ model with order parameter $r = 10$.

    \item Third coefficient of the autoregressive $AR(r)$ model with order parameter $r = 10$.

    \item Sixth coefficient of the autoregressive $AR(r)$ model with order parameter $r = 10$.

    \item Mean over the absolute differences between subsequent values ($\Lambda$): This feature describes the average rate of change with a time series and was computed using the following equation.
    \begin{equation}
        \Lambda=\frac{1}{n-1} \sum_{i=1, \ldots, n-1}\left|x_{i+1}-x_i\right|
    \end{equation}
    This feature becomes a measure of the magnitude of fluctuations existing within the time series. A high value of $\Lambda$ denotes significant variations between successive data points. A low value of $\Lambda$ denotes a more stable and less volatile time series.

    \item Percentage of reoccurring values to all values ($\Delta$): This feature is a ratio of the number of values that occur more than once within a time series to the total number of unique values. Physically, this feature quantifies the degree of regularity in the time series data. In other words, this feature could provide insights into the temporal stability or periodicity of the flow. A higher $\Delta$ value would indicate a more regular and predictable flow behaviour. Whereas, a lower $\Delta$ value would indicate irregularity or variability in the flow behaviour. Since we are dealing with turbulent flows, this feature is expected to have fairly low values, which may vary among different orifice geometries.

    \item Fifth coefficient of spectral welch density ($\xi$): This feature estimates the cross-power spectral density of a time series $x$, at different frequencies using Welch's method. For this feature extraction, the time series was initially shifted from the time domain to the frequency domain. The fifth coefficient of the power spectral density characterizes the power distribution at the fifth frequency component. In the context of fluid dynamics, these features may capture periodicity within the flow. When external perturbations such as vortex shedding past orifice plates exist, the power spectral density corresponding to the shedding frequency would be non-zero. Different orifice geometries could induce different levels of vortex shedding locally at the measurement points.

    \item Second coefficient of spectral welch density ($\xi$): Like the previous feature, the second coefficient of the power spectral density characterizes the power distribution at the second frequency component.

    \item The value of partial autocorrelation function at a lag of $3$: The partial autocorrelation is a statistical measure that quantifies the linear relationship between a time series variable and its lagged values. In the context of our pipe flow, partial autocorrelation can provide insights into the flow velocity's temporal dependence and correlation structure. This means that this feature can be useful in understanding the persistence or memory of the signal. It suggests that a strong linear relationship between the current flow state and its state $3$ time steps ago has been important for classifying flow past orifices. In this analysis, a `time step' corresponds to the original sampling rate of $10\mathrm{kHz}$. Therefore, a lag of $3$ time steps will signify a duration of $0.3$ milliseconds.

    \item The value of partial autocorrelation function at a lag of $6$.

    \item The value of partial autocorrelation function at a lag of $8$.

    \item Kurtosis of the velocity time series calculated with the adjusted Fisher-Pearson standardized moment coefficient, $g2$: We know that Kurtosis is a higher-order statistical attribute of velocity signals. The heaviness of the tails of the probability density functions of normalized time series could be distinct for the signals corresponding to each orifice geometry. This feature will help us assess the degree of deviation from the Gaussian distribution and provide evidence of the skewed behaviour of the time series.
    
\end{enumerate}
A general examination of the list of discriminatory features suggests that microscale autocorrelation coefficients and other such quantified parameters contain sufficient information to differentiate the jet arising out of one orifice to another. It was also observed that the actual ordering of the important features can change with $Re$, but the characteristics of the list are largely unchanged.

\section*{Conclusion}

The analysis presented in this manuscript focused on understanding the flow physics in the near field and downstream of different orifice geometries using velocity time series measurements. The performance of the orifice identification algorithm was evaluated on a dataset of $25$ different orifice geometries. Remarkably, for the recorded dataset, the algorithm identified the geometries with $100\%$ accuracy and $100\%$ precision. This implicitly meant that the orifice confirmation worked perfectly, which was evaluated and observed to achieve a true confirmation rate of $100\%$. The dataset showed similar performance for both types of identification algorithms discussed in this thesis, namely single random forest and class binarisation approach. The clustering of orifice geometries based on their flow characteristics was explored. Various clusters or classes of orifice geometries were visualised using the outcomes from the class binarisation approach.

We effectively sample the three-dimensional transient velocity profile (3-dimensional fluctuations of velocity profile) downstream of the orifices. The features of such time series have been shown to be sufficient for machine learning classifiers to identify the flow uniquely and, in turn, identify the geometry of the obstruction. The repeatability and robustness of the experiments were tested for three geometries: square, circle and equilateral triangle orifice geometries under different system alignments. The model achieved $100\%$ prediction accuracy for all three geometries for different probe alignments with respect to the pipe axis. It also performed well for various combinations of the anemometer module and hot film probe. This implies that the proposed method is robust to both alignment and instrumentation variations. The impact of various system Reynolds numbers on the performance of a binary classifier trained to classify square and circle orifice geometries was investigated. From those observations, it was concluded that training the model for a Reynolds number of $2.37\times 10^4$ or $2.77\times 10^4$ is ideal since it can be used across a wide range of Reynolds numbers.

The performance of the random forest model was investigated for varying distances downstream for the classification of square- and circle-shaped orifices. As expected, the model accuracy was observed to decrease downstream, with $100\%$ at $z=0.6D$, $97\%$ at $z=1.9D$ and $78\%$ at $z=4.2D$. The observations from this study implied that the information on the orifice geometry is partially preserved in the flow field downstream, with its strength remaining high until a distance nearly equal to double the orifice diameter. As expected, this information would be gradually lost beyond this distance, owing to the nature of a fully developed jet.

Important features extracted from the random forest models were analysed to gain insights into the underlying flow physics responsible for classification. These features, each interpreted in the context of flow behaviour, provide valuable information about the temporal dependencies and complexity of the flow past orifices of different geometries. Overall, the outcomes from this study suggest that upstream geometry will leave an indelible fluid mechanic signature on the flow field.

While the proposed study primarily focuses on identifying the geometry of orifices in fluid flow systems, it is possible that the techniques and methods developed as a result of this work find applications in the medical field, especially pulmonary studies. Hot wire anemometry is a well-established technique in fluid dynamics and has been used in a variety of medical research, including the measurement of volumetric flow of exhaled breath. The techniques and methods developed in this manuscript could potentially be adapted to measure and analyze the airflow in the respiratory system, including the identification of airway obstructions or other abnormalities. However, further research would be necessary to explore the potential applications of the proposed techniques in the medical field.

\subsection*{Future scope of the orifice identification dataset}

The dataset obtained in this manuscript potentially opens up several interesting directions for future research in the area of machine learning applications in experimental fluid dynamics. In order to make advancements in the area of flow predictions and its control, sophisticated algorithms are needed that could reconstruct the flow passage and obstruction geometries using simplified measurement techniques. With the available insights from this study, the next steps in problem formulation would be as follows:

\subsubsection*{Feature generation} 

A forward problem can be formulated in which a deep-learning model generates a set of time series features given an image (for example, a $50 \times 50$ \textit{.jpg} or \textit{.png} file) of the orifice geometry. The output features from this approach can be compared with the set of pre-extracted features for evaluating the model. The feature set can then be used to reconstruct the time series itself, and this would make this algorithm an effective ``Neural network hot wire anemometer''.

\subsubsection*{Generative problem}

As a generative version of this problem, a deep-learning model could be built that could generate the image of the orifice which would yield from a given set of input time series features. Consider a trained model based on $20$ orifice geometries. When a time series is given as input, especially one that is outside the set from these $20$ geometries, the model should attempt to construct the image of the orifice. This algorithm can be considered as an alternative version of a large-language model specialising in fluid mechanics.

These studies will shed light on the intricate relationship between orifice shape and flow behaviour, highlighting the significance of geometrical variations in influencing flow patterns and turbulence within flow domains.

\acknow{The authors acknowledge the SENAI of Indian Institute of Technology Madras (IITM) for providing the required high-performance computing resources; the NCCRD, IITM, for providing the hot-wire anemometer setup and the calibration facility; and Dr. R I Sujith, Professor, IITM for sharing experimental facilities.}

\showacknow{} 

\bibsplit[10]


\begin{thebibliography}{10}

\bibitem{Reynolds1883}
O Reynolds, An experimental investigation of the circumstances which determine whether the motion of water shall be direct or sinuous, and of the law of resistance in parallel channels.
\newblock {\em\protect\JournalTitle{Philosophical Transactions of the Royal Society of London}} \textbf{174}, 935--982 (1883).

\bibitem{Curtet1958}
R Curtet, Confined jets and recirculation phenomena with cold air.
\newblock {\em\protect\JournalTitle{Combustion and Flame}} \textbf{2}, 383–411 (1958).

\bibitem{Rajaratnam1976confinedjets}
N Rajaratnam, ed., Chapter 8 confined jets in {\em Turbulent Jets}, Developments in Water Science.
\newblock (Elsevier) Vol.{}~5, pp. 148--183 (1976).

\bibitem{Exley1971}
JT Exley, JA Brighton, Flow separation and reattachment in confined jet mixing.
\newblock {\em\protect\JournalTitle{Journal of Basic Engineering}} \textbf{93}, 192–198 (1971).

\bibitem{Kandakure2008}
M Kandakure, V Patkar, A Patwardhan, Characteristics of turbulent confined jets.
\newblock {\em\protect\JournalTitle{Chemical Engineering and Processing: Process Intensification}} \textbf{47}, 1234–1245 (2008).

\bibitem{Muller1999}
M Muller, S.~Milano, P Koumoutsakos, Application of machine learning algorithms to flow modeling and optimization (1999).

\bibitem{Zhang2015}
ZJ Zhang, K Duraisamy, Machine learning methods for data-driven turbulence modeling in {\em 22nd AIAA Computational Fluid Dynamics Conference}.
\newblock (American Institute of Aeronautics and Astronautics), (2015).

\bibitem{Beck2021}
A Beck, M Kurz, A perspective on machine learning methods in turbulence modeling.
\newblock {\em\protect\JournalTitle{GAMM-Mitteilungen}} \textbf{44} (2021).

\bibitem{Kutz2017}
JN Kutz, Deep learning in fluid dynamics.
\newblock {\em\protect\JournalTitle{Journal of Fluid Mechanics}} \textbf{814}, 1–4 (2017).

\bibitem{Karniadakis2021}
GE Karniadakis, et~al., Physics-informed machine learning.
\newblock {\em\protect\JournalTitle{Nature Reviews Physics}} \textbf{3}, 422–440 (2021).

\bibitem{Schlkopf2018}
B Sch\"{o}lkopf, AJ Smola, {\em Learning with Kernels: Support Vector Machines, Regularization, Optimization, and Beyond}.
\newblock (The MIT Press), (2018).

\bibitem{Breiman2001}
L Breiman, Random forests.
\newblock {\em\protect\JournalTitle{Machine Learning}} \textbf{45}, 5--32 (2001).

\bibitem{Zhou2020}
Y Zhou, D Fan, B Zhang, R Li, BR Noack, Artificial intelligence control of a turbulent jet.
\newblock {\em\protect\JournalTitle{Journal of Fluid Mechanics}} \textbf{897} (2020).

\bibitem{Li2022}
Y Li, J Chang, C Kong, W Bao, Recent progress of machine learning in flow modeling and active flow control.
\newblock {\em\protect\JournalTitle{Chinese Journal of Aeronautics}} \textbf{35}, 14–44 (2022).

\bibitem{Fan2020}
D Fan, L Yang, Z Wang, MS Triantafyllou, GE Karniadakis, Reinforcement learning for bluff body active flow control in experiments and simulations.
\newblock {\em\protect\JournalTitle{Proceedings of the National Academy of Sciences}} \textbf{117}, 26091–26098 (2020).

\bibitem{Brunton2020b}
SL Brunton, MS Hemati, K Taira, Special issue on machine learning and data-driven methods in fluid dynamics.
\newblock {\em\protect\JournalTitle{Theoretical and Computational Fluid Dynamics}} \textbf{34}, 333–337 (2020).

\bibitem{Brunton2020a}
SL Brunton, BR Noack, P Koumoutsakos, Machine learning for fluid mechanics.
\newblock {\em\protect\JournalTitle{Annual Review of Fluid Mechanics}} \textbf{52}, 477–508 (2020).

\bibitem{Brunton2021}
SL Brunton, Applying machine learning to study fluid mechanics.
\newblock {\em\protect\JournalTitle{Acta Mechanica Sinica}} \textbf{37}, 1718–1726 (2021).

\bibitem{Li2020turbulentregion}
B Li, et~al., Using machine learning to detect the turbulent region in flow past a circular cylinder.
\newblock {\em\protect\JournalTitle{Journal of Fluid Mechanics}} \textbf{905}, A10 (2020).

\bibitem{Fukami2019}
K Fukami, K Fukagata, K Taira, Super-resolution reconstruction of turbulent flows with machine learning.
\newblock {\em\protect\JournalTitle{Journal of Fluid Mechanics}} \textbf{870}, 106–120 (2019).

\bibitem{Fukami2020}
K Fukami, K Fukagata, K Taira, Machine-learning-based spatio-temporal super resolution reconstruction of turbulent flows.
\newblock {\em\protect\JournalTitle{Journal of Fluid Mechanics}} \textbf{909} (2020).

\bibitem{Fukami2023}
K Fukami, K Fukagata, K Taira, Super-resolution analysis via machine learning: a survey for fluid flows.
\newblock {\em\protect\JournalTitle{Theoretical and Computational Fluid Dynamics}} \textbf{37}, 421–444 (2023).

\bibitem{Faouzi2022}
J Faouzi, {Time Series Classification: A Review of Algorithms and Implementations} in {\em Machine Learning (Emerging Trends and Applications)}, ed.{} K Kotecha.
\newblock (Proud Pen), (2022).

\bibitem{Acharya2021}
AS Acharya, S Deevi, K Dhivyaraja, AK Tangirala, MV Panchagnula, Spatio-temporal microstructure of sprays: data science-based analysis and modelling.
\newblock {\em\protect\JournalTitle{Journal of Fluid Mechanics}} \textbf{912}, A19 (2021).

\bibitem{Godavarthi2019}
V Godavarthi, K Dhivyaraja, RI Sujith, MV Panchagnula, Analysis and classification of droplet characteristics from atomizers using multifractal analysis.
\newblock {\em\protect\JournalTitle{Scientific Reports}} \textbf{9} (2019).

\bibitem{Oriona2021}
{\'A} L{\'o}pez-Oriona, JA Vilar, F4: An all-purpose tool for multivariate time series classification.
\newblock {\em\protect\JournalTitle{Mathematics}} \textbf{9} (2021).

\bibitem{nun2015fats}
I Nun, et~al., Fats: Feature analysis for time series (2015).

\bibitem{CHRIST2018}
M Christ, N Braun, J Neuffer, AW Kempa-Liehr, Time series feature extraction on basis of scalable hypothesis tests (tsfresh – a python package).
\newblock {\em\protect\JournalTitle{Neurocomputing}} \textbf{307}, 72--77 (2018).

\bibitem{Fulcher2017}
BD Fulcher, NS Jones, hctsa: A computational framework for automated time-series phenotyping using massive feature extraction.
\newblock {\em\protect\JournalTitle{Cell Systems}} \textbf{5}, 527--531.e3 (2017).

\bibitem{BARANDAS2020}
M Barandas, et~al., Tsfel: Time series feature extraction library.
\newblock {\em\protect\JournalTitle{SoftwareX}} \textbf{11}, 100456 (2020).

\end{thebibliography}

\end{document}